\documentstyle[epsfig,aps,pra]{revtex}
\tighten
\begin{document}
\title{Jost Function for Singular Potentials}
\author{S. A. Sofianos, S. A. Rakityansky, and S. E. Massen\thanks{
Permanent address: Department of Theoretical Physics,
University of Thessaloniki, Thessaloniki, 54006, Greece}}
\address{Physics Department, University of South Africa,
 	         P.O.Box 392, Pretoria 0003, South Africa}
\date{\today}
\maketitle
\begin{abstract}
An exact method for direct calculation of the Jost function and Jost
solutions for a repulsive singular potential is presented. Within this method
the Schr\"odinger equation is replaced  by an equivalent system of linear
first--order differential equations, which after complex rotation,
can easily be solved numerically. The Jost function can be obtained 
to any desired accuracy for all complex momenta of physical interest, 
including the spectral points corresponding to bound
and resonant states. The method can  also be used in the complex
angular--momentum plane to calculate the Regge trajectories. The 
effectiveness of the method is demonstrated using the Lennard--Jones 
(12,6) potential. The spectral properties of the realistic inter--atomic 
$^4$He--$^4$He potentials HFDHE2 and HFD-B of Aziz and collaborators 
are also investigated.
\end{abstract}
\vspace{0.5cm}
\section{Introduction}
A new method for locating potential resonances and Regge trajectories,
based on  direct calculation of the Jost function in the complex
$k$--plane, has  recently been developed  \cite{nuovocim,exact,nnn,coupled}.
Within this method,  the bound, resonant, and scattering states can be found
by calculating  the Jost solutions and the Jost function on  the  appropriate
domain of the $k$--plane. The bound and resonant state energies, for example,
can be found by locating  the  zeros of the Jost function on the positive
imaginary axis and in the fourth quadrant respectively. At the same time,
as a by-product of the Jost function  calculation, one gets the physical
wave function that has the correct asymptotic  behavior.\\

The method, in the form developed in Refs. \cite{nuovocim,exact,nnn,coupled},
cannot be directly applied to potentials which are more singular 
than $1/r^2$ at the origin. A significant number of practical problems, 
however, where the Jost function could be very useful, involves such 
potentials. For example, inter-atomic and inter-molecular forces at short
distances are strongly repulsive due to the overlap of the electron clouds,
and thus they are usually represented by repulsive singular potentials
such as the Lennard--Jones(12,6) one which has $\sim 1/r^{12}$ behavior. 
It is, therefore, desirable  to extend the Jost function method for 
potentials of this kind.\\

It is well-known that attractive singular potentials do not admit physically
meaningful solutions with the usual boundary conditions \cite{frank}.
All solutions of the Schr\"odinger equation with such  potentials vanish at
the origin and there is no apparent way to determine the arbitrary phase
factor between them. In contrast, repulsive singular potentials do not
pose any problem regarding mathematical uniqueness or physical
interpretation. However, the integration of the Schr\"odinger equation
as well as of the relevant equations for the Jost function, have inherent
difficulties resulting from the fact that the singularity
of the  potential makes the  $r=0$ an irregular singular point
of the equation.  In particular, the regular solution cannot be
defined by universal boundary conditions independent of the potential.\\

This drawback,  stemming from the  extremely strong repulsion near the origin,
can fortunately be tackled by the  WKB approximation  which provides the
correct radial behavior of the wave function in the neighborhood
of the point $r=0$ \cite{landau}. Therefore, starting with the WKB
boundary conditions, one  can find the regular solution by integrating the
Schr\"odinger equation from $r=0$ to some intermediate point $r=r_{\rm int}$.
Then  the equations for the Jost function can be integrated,
 from $r_{\rm int}$  outwards as for the nonsingular potential using
the boundary conditions at  $r=r_{\rm int}$ which are expressed in
terms of the regular solution and its first derivative.\\

The paper is organized as follows: In Sec. II our formalism  is
presented and  is  tested in Sec. III  using an example known in the
literature; in Sec. IV the method is applied to realistic
interaction between helium atoms. Our conclusions  are drawn
in Sec. V.
\section{Theory}
\subsection{Basic equations and definitions}
There are three different types of physical problems associated with the
Schr\"odinger equation ($2\mu/\hbar^2=1$),
\begin{equation}
\label{schr}
	       \left[\partial^2_r+k^2-{\ell(\ell+1)}/{r^2}\right]
	       u_{\ell}(k,r)=V(r)u_{\ell}(k,r)\ ,
\end{equation}
namely,  bound, scattering, and resonant state problems. They differ in the
boundary conditions imposed on the wave function at large distances.
Alternatively, a solution can be prescribed by the boundary conditions
at the origin. In the case of regular potentials obeying  the condition
\begin{equation}
\label{regconV}
	      \lim_{r\to 0}r^2 V(r) = 0 \ ,
\end{equation}
the solution $\phi_\ell(k,r)$ which vanishes near $r=0$ exactly like the
Riccati--Bessel function,
\begin{equation}
\label{regconF}
	      \lim_{r\to 0} \phi_\ell(k,r)/j_\ell(kr) = 1\ ,
\end{equation}
is called the {\it regular} solution. Since all physical solutions are
regular at the origin, they differ from $\phi_\ell(k,r)$ only by a
normalization constant. Therefore, if the function $\phi_\ell(k,r)$  
can be calculated at all real and complex momenta $k$, one can have, 
in principle, all solutions of physical interest in a most general form.
For example, the calculation of the scattering solutions on the real 
$k$-axis is simply a matter of finding the proper normalization for
$\phi_\ell(k,r)$ because the regular solution has the correct
behavior at large $r$ for any $k>0$. In contrast, to find the bound
and resonant states where the  $k$ is  complex, one must ensure
that the function $\phi_\ell(k,r)$ has the proper physical asymptotic 
behavior which exist  only at certain  points on the  $k$--plane. 
These spectral points can be found by many different ways, but perhaps 
the most convenient way to find them is by locating the zeros of the 
Jost function.\\

For any complex $k$, the regular solution at large distances  can be
expressed as a linear combination of the Riccati--Hankel functions
$h_\ell^{(\pm)}(kr)$,
\begin{equation}
\label{assF}
      \phi_\ell(k,r)\mathop{\longrightarrow}\limits_{r\to\infty}
      \frac12\left[h_\ell^{(+)}(kr)f^*_\ell(k^*)+
      h_\ell^{(-)}(kr)f_\ell(k)\right]\ ,
\end{equation}
where the $r$-independent but momentum--dependent coefficient $f_\ell(k)$ is
the Jost function. From the asymptotic form (\ref{assF}) it is clear that
the zeros of $f_\ell(k)$ on the positive imaginary axis of the complex
$k$--plane correspond to bound states while those in the
fourth quadrant to resonances.\\

In order to find $f_\ell(k)$, we look for the regular solution on the whole
interval $[0,\infty)$ in the form
\begin{equation}
\label{ansatz}
	\phi_\ell(k,r)=\frac12\left[
	h_{\ell}^{(+)}(kr)F_\ell^{(+)}(k,r)+
	h_{\ell}^{(-)}(kr)F_\ell^{(-)}(k,r)\right]\ ,
\end{equation}
where the new unknown functions $F_\ell^{(\pm)}(k,r)$ are subjected to
the additional condition
\begin{equation}
\label{lagrange}
	h_\ell^{(+)}(kr)\partial_rF_\ell^{(+)}(k,r)+
	h_\ell^{(-)}(kr)\partial_rF_\ell^{(-)}(k,r)=0\ .
\end{equation}
Eq. (\ref{schr}) is then  transformed into an equivalent system of
first order equations
\begin{equation}
\label{fpmeq}
\partial_rF_\ell^{(\pm)}(k,r)=\pm\displaystyle{
	  \frac{h_\ell^{(\mp)}(kr)}{2ik}
	  V(r)\left[
	  h_{\ell}^{(+)}(kr)F_\ell^{(+)}(k,r)+
	  h_{\ell}^{(-)}(kr)F_\ell^{(-)}(k,r)\right]}\ .
\end{equation}
In Refs.\cite{nuovocim,coupled} it was shown that at large distances
$F_\ell^{(-)}(k,r)$ coincides with the Jost function,
\begin{equation}
\label{flim}
	  \lim_{r\to\infty}F_\ell^{(-)}(k,r)=f_\ell(k)\ ,
\end{equation}
but this limit only exists when
\begin{equation}
\label{limcond}
	  {\rm Im\,}kr\ge 0\ .
\end{equation}
If $r$ is real, the condition (\ref{limcond}) is only satisfied for bound
and scattering states but not for resonances. To calculate $f_\ell(k)$
we, therefore, make a complex rotation of the coordinate in Eqs. 
(\ref{fpmeq}),  in the first  quadrant 
\begin{equation}
\label{rot}
	r=x\exp(i\theta)\, ,\qquad x\ge 0\, ,
		\qquad 0\le\theta<\frac{\pi}{2}\ ,
\end{equation}
with a sufficiently large $\theta$ (see Refs.
\cite{nuovocim,exact,nnn,coupled} for more details). Such a rotation is only
possible if the potential is an analytic function of $r$ and tends to zero
when $x\to\infty$ for the chosen angle $\theta$.
%
\subsection{Boundary conditions}
In the case of regular potentials the boundary conditions for
Eqs. (\ref{fpmeq}) are very simple,
\begin{equation}
\label{fcondr}
	  F_\ell^{(\pm)}(k,0)=1\ .
\end{equation}
They follow immediately from (\ref{regconF}), (\ref{ansatz}), and the fact
that
$$
	\frac12\left[h_{\ell}^{(+)}(kr)+
	h_{\ell}^{(-)}(kr)\right]=j_\ell(kr)\ .
$$
Going  over to singular potentials, Eq. (\ref{regconF}) does not hold
anymore. Due to the extremely strong repulsion, the regular
solution vanishes much faster than $j_\ell(kr)$ when $r\to0$.
In fact,  it vanishes exponentially \cite{newton} and therefore the
conditions (\ref{fcondr}) must be modified accordingly. 
In order to find the exact behavior of the regular solution near 
the origin we apply the familiar semi-classical WKB method.
Though  the strong repulsion makes things rather complicated, it has 
the  advantage that the criterion of the applicability of the WKB 
approximation is satisfied when $r\to 0$.
Indeed, the WKB method works well  when the local wavelength
$\lambda$  varies slowly, {\em i.e.}
\begin{equation}
\label{wkbcond}
		 |{\rm d}\lambda/{\rm d}r|\ll 1\ .
\end{equation}
It can be shown \cite{frank} that this derivative is given by
\begin{equation}
\label{crit}
     |{\rm d}\lambda/{\rm d}r|=\frac12
     \left|\frac{{\rm d}V(r)}{{\rm d}r}[k^2-V(r)]^{-3/2}\right|\ .
\end{equation}
Assuming that $V(r)$ approaches its singularity near $r=0$ monotonically,
we can find an $r_{\rm min}$ that for all $r<r_{\rm min}$ the momentum  
in (\ref{crit}) is negligible, {\em i.e.} we may write
\begin{equation}
\label{crit1}
      |{\rm d}\lambda/{\rm d}r|~\mathop{\longrightarrow}\limits_{r\to
      0}~\frac12\left|\frac{{\rm d}V(r)}{{\rm d}r}[V(r)]^{-3/2}\right|\ .
\end{equation}
When $r\to 0$, the right hand side of Eq. (\ref{crit1})
for  usual singular potentials tends to zero. For example, if
$$
	  V(r)~\mathop{\longrightarrow}\limits_{r\to0}~g/r^n\ ,
$$
the condition (\ref{wkbcond}) is always satisfied for $n>2$,
$$
      |{\rm d}\lambda/{\rm d}r|~\mathop{\longrightarrow}\limits_{r\to
      0}~\frac{nr^{\frac12n-1}}{2\sqrt{g}}\longrightarrow 0\ ,
      \qquad{\rm if}\ n>2\ .
$$
Therefore, assuming that the necessary condition (\ref{wkbcond}) is
fulfilled and choosing a small enough $r_{\rm min}$, we can express
the regular solution on the interval $[0,r_{\rm min}]$ using the
WKB approximation (see, for example, Ref.\cite{connor}), viz.
\begin{equation}
\label{wkbfun}
       \phi_\ell(k,r)=\frac{1}{\sqrt{p(r)}}
       \exp\left[i\int_r^a p(\rho){\rm d}\rho\right]\ ,
       \qquad r\in[0,r_{\rm min}]\ ,
\end{equation}
where the classical momentum $p(r)$ is defined by
\begin{equation}
\label{classmom}
       p(r)\equiv\sqrt{k^2-V(r)-(\ell+\frac12)^2/r^2}
\end{equation}
and the upper limit $a$ in the integral is an arbitrary value
$a>r_{\rm min}$. Usually $a$ is taken to be the inner turning point
\cite{connor}, but it is obvious from Eq. (\ref{wkbfun}) that an additional
integration from $a$ to the turning point can only change the overall
normalization of the solution which is not our concern at the moment.
Thus, Eq. (\ref{wkbfun}) together with the derivative
\begin{equation}
\label{wkbder}
       \partial_r\phi_\ell(k,r)=\left\{
       \frac{\displaystyle   \frac{{\rm d}V(r)}{{\rm d}r}
       -2\left(\ell+\frac12\right)^2r^{-3} }
       {  4[p(r)]^{5/2} }
       -i\sqrt{p(r)}   \right\}
       \exp\left[i\int_r^a p(\rho){\rm d}\rho\right]\ ,
       \qquad r\in(0,r_{\rm min}]
\end{equation}
can be used as  boundary conditions for the regular solution of the
Schr\"odinger equation at any point in the interval $(0,r_{\rm min}]$.
To obtain the corresponding boundary conditions for the functions
$F_\ell^{(\pm)}(k,r)$, we need to express them in terms of $\phi_\ell(k,r)$
and $\partial_r\phi_\ell(k,r)$. For this we can use Eq. (\ref{ansatz})
together with relation
\begin{equation}
\label{lagr}
      \partial_r\phi_\ell(k,r)=\frac{1}{2}\left[F_\ell^{(+)}(k,r)
      \partial_r h_\ell^{(+)}(kr)+ F_\ell^{(-)}(k,r)
      \partial_r h_\ell^{(-)}(kr)\right]\ ,
\end{equation}
which follows from (\ref{lagrange}). From (\ref{ansatz}) and
(\ref{lagr}) we find that
\begin{equation}
\label{match}
      F_\ell^{(\pm)}(k,r)=\pm \frac{i}{k}\left[\phi_\ell(k,r)
      \partial_r h_\ell^{(\mp)}(kr) - h_\ell^{(\mp)}(kr)
      \partial_r\phi_\ell(k,r)\right]
\end{equation}
which is valid for any $r\in[0,\infty)$. Therefore Eqs. (\ref{match}) taken
at some point $r<r_{\rm min}$ with $\phi_\ell(k,r)$ and
$\partial_r\phi_\ell(k,r)$ given by (\ref{wkbfun}) and (\ref{wkbder}),
provide us the  boundary conditions, required in Eqs. (\ref{fpmeq}),
for singular potentials. It can easily be checked (by using 
$j_\ell(kr)$ for the regular solution near $r=0$)  that Eq. (\ref{match}) 
gives the correct boundary conditions  for regular potentials as well, \\

Alternatively to impose the boundary conditions on the functions
$F_\ell^{(\pm)}(k,r)$ near the origin, one can simply solve the Schr\"odinger
equation from a small $r$ up to some intermediate point $b$ where, using
(\ref{match}), the $F_\ell^{(\pm)}(k,b)$ can be obtained and propagated
further on by integrating equations (\ref{fpmeq}).
%
\subsection{Integration path}
The use of more complicated boundary conditions at $r=0$ does not change
the condition (\ref{limcond}) for the  existence of the limit (\ref{flim}).
Indeed, in deriving this condition we  used only the behavior
of the potential and the Riccati--Hankel functions at large distances
\cite{nuovocim,coupled}. Therefore, the Jost function for a singular
potential can also be calculated by evaluating the function
$F_\ell^{(-)}(k,r)$ at a large $r$. When we are dealing with resonances,
{\rm i.e.} working in the fourth quadrant of the $k$--plane, we need to
integrate Eqs. (\ref{fpmeq}) along the turned ray (\ref{rot}).\\

\begin{center}
\begin{figure}
\centerline{\epsfig{file=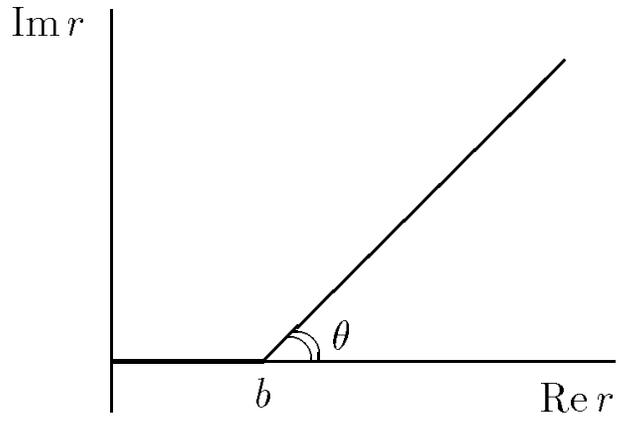,height=6.65092cm,width=10cm}}
\caption{
Deformed contour for integration of the differential equations}
\label{contour}
\end{figure}
\end{center}

As can be seen from the WKB boundary conditions (\ref{wkbfun}), the use of 
a complex $r$ near the origin, makes $\phi_\ell(k,r)$   oscillatory 
from the outset. Although this does not formally cause any problem, 
in numerical calculations such oscillations may reduce the accuracy. 
To avoid this  we solve Eqs. (\ref{fpmeq}) from a small
$r_{\rm min}$ to some intermediate point $b$ along the real axis
and then perform the  complex rotation,
\begin{equation}
\label{drot}
      r=b+x\exp(i\theta)\, ,\quad x\in[0,\infty)\, ,
		\quad 0\le\theta<\frac{\pi}{2}\ ,
\end{equation}
as is shown in Fig. \ref{contour}. Therefore, on the interval 
$[r_{\rm min},b]$ we can use Eqs. (\ref{fpmeq})
as they are, while beyond the point $r=b$ these equations are transformed to
\begin{eqnarray}
\nonumber
       \partial_xF_\ell^{(\pm)}(k,b+xe^{i\theta})=&\pm&\displaystyle{
	  \frac{e^{i\theta}h_\ell^{(\mp)}(kb+kxe^{i\theta})}{2ik}
	  V(b+xe^{i\theta})}\left[h_{\ell}^{(+)}(kb+kxe^{i\theta})
	  F_\ell^{(+)}(k,b+xe^{i\theta})\right.\\
\label{fpmeq1}
&&\\
\nonumber
	  &+& \left. h_{\ell}^{(-)}(kb+kxe^{i\theta})
	  F_\ell^{(-)}(k,b+xe^{i\theta})\right]\ .
\end{eqnarray}
Though the complex transformation (\ref{drot}) is different from
(\ref{rot}), the proof of the existence of the limit (\ref{flim}) given in
the Appendix A.2 of Ref. \cite{coupled} remains applicable here. 
Indeed, that proof was based on the fact that for ${\rm Im\,}kr>0 $ the
Riccati--Hankel function $h_{\ell}^{(+)}(kr)$  decays exponentially
at large $|r|$, and thus the derivative $\partial_rF^{(-)}(k,r)$
vanishes there and  the function $F^{(-)}(k,r)$ becomes a constant.
Under the transformation (\ref{drot}) the asymptotic behavior of the
Riccati--Hankel function,
\begin{equation}
\label{richaninf}
	h_\ell^{(+)}(kr)\mathop{\longrightarrow}\limits_{r\to\infty}
	-i\exp\left[i(kr-{\ell\pi/2})\right]\ ,
\end{equation}
has only an additional $r$--independent phase  factor $\exp(ikb)$ which
does not affect the proof.\\

From the above, it is clear that we can identify  the Jost function
$f_\ell(k)$ as the value of $F_\ell^{(-)}(k,b+xe^{i\theta})$ at a
sufficiently large $x$ beyond which this function is practically  constant.
In the bound and scattering state domain, where ${\rm Im\,}k\ge 0 $,
one  can choose any rotation angle $\theta$ allowed by the potential, 
including $\theta=0$. In the resonance domain, however, where
$$
	       k=|k|\exp(-i\varphi)\ , \qquad \varphi>0\ ,
$$
the rotation angle $\theta$ must be greater or equal to $\varphi$. If
the condition $\theta \ge \varphi$ is fulfilled, the value of the limit
(\ref{flim}) does not depend on the choice of $\theta$. This provides us
with a reliable way to check the stability and accuracy of the calculations 
by comparing the results for $f_\ell(k)$ obtained with two different
values of $\theta$.\\

From Eq.  (\ref{richaninf}) it is clear that the angular momentum appears
only in the phase factor of the asymptotic behavior of the Riccati--Hankel
functions and hence of the regular solution. Therefore, the use of any 
complex $\ell$ cannot change the domain of the $k$--plane where the limit
(\ref{flim}) exists. This means that the Jost function can be calculated, 
for any complex angular momentum, using the same equations. Moreover, 
when looking for  the Regge poles in the $\ell$--plane, the complex 
rotation  is not necessary  because these poles correspond to real
energies. Locating Regge poles as zeros of the Jost function in the
complex $\ell$--plane is easier than by calculating them via the
$S$--matrix using three integration paths (in the $r$--plane)
as  suggested in Ref. \cite{sukumar}.
\subsection{Jost solutions}
\label{jsol}
By storing  the values of $F_\ell^{(\pm)}(k,r)$ on the integration grid 
one  can also obtain the regular solution in the form (\ref{ansatz}) 
on the  interval $[r_{\rm min},r_{\rm max}]$. It is noted
that the  use  of  the Riccati--Hankel functions in (\ref{ansatz})  
guarantees the correct (in fact exact) asymptotic behavior of the 
wave function.\\

The regular solution thus obtained consists of two terms:
\begin{eqnarray}
\nonumber
     \frac12 h_\ell^{(+)}(kr)F_\ell^{(+)}(k,r)\
     \mathop{\longrightarrow}_{r\to\infty}\
     &-&\frac{i}{2}\exp\left[+i(kr-\ell\pi/2)\right]f_\ell^*(k^*)\ ,\\
\nonumber
&&\\
\nonumber
     \frac12 h_\ell^{(-)}(kr)F_\ell^{(-)}(k,r)\
     \mathop{\longrightarrow}_{r\to\infty}\
     &+&\frac{i}{2}\exp\left[-i(kr-\ell\pi/2)\right]f_\ell(k)\ .
\end{eqnarray}
Asymptotically they behave like $\sim {\rm e}^{\pm ikr}$ and thus at long
distances they are proportional to the commonly used {\it Jost solutions}
${\rm f}_\ell^{(\pm)}(k,r)$ for which
\begin{equation}
\label{defjsol}
     {\rm f}_\ell^{(\pm)}(k,r)\
     \mathop{\longrightarrow}_{r\to\infty}\ h_\ell^{(\pm)}(kr)\ .
\end{equation}
In practice, the Jost solutions can be calculated, via (\ref{ansatz}), 
by integrating Eqs. (\ref{fpmeq}) inwards from a sufficiently 
large $r_{\rm max}$ with the boundary conditions
$$
\left[
     \begin{array}{c}
     F_\ell^{(+)}(k,r_{\rm max})\\
     F_\ell^{(-)}(k,r_{\rm max})
     \end{array}
\right]=
\left[
     \begin{array}{c}
     2\\
     0
     \end{array}
\right]\ ,\qquad \mbox{for\quad  ${\rm f}_\ell^{(+)}(k,r)$}\ ,
$$
$$
\left[
     \begin{array}{c}
     F_\ell^{(+)}(k,r_{\rm max})\\
     F_\ell^{(-)}(k,r_{\rm max})
     \end{array}
\right]=
\left[
     \begin{array}{c}
     0\\
     2
     \end{array}
\right]\ ,\qquad \mbox{for\quad  ${\rm f}_\ell^{(-)}(k,r)$}\ ,
$$
which obviously comply with the definition (\ref{defjsol}). The advantage of
such an approach is that  at large $r$ all the oscillations of
${\rm f}_\ell^{(\pm)}(k,r)$ are described  exactly by the Riccati--Hankel
functions while the functions $F_\ell^{(\pm)}(k,r)$ are smooth.
\section{Lennard--Jones potential}
In order to evaluate the accuracy and efficiency of our method
we apply it to the Lennard--Jones potential
\begin{equation}
\label{LJpot}
	    V(r)=D\left[\left(\frac{d}{r}\right)^{12}-
	    2\left(\frac{d}{r}\right)^{6}\right]\ .
\end{equation}
which is   well--known in atomic  and molecular physics.
Combined with a rotational barrier, this potential supports narrow as
well as broad resonant states (see, for example, Ref.\cite{connor}).
To locate them, any method employed must be pushed to the extreme, thus
exhibiting  its advantages and drawbacks.\\

To be able to compare our results with other calculations, we chose the
parameters in (\ref{LJpot}) to be the same as those  used in
Refs. \cite{connor,weon}, namely, $d=3.56$\,\AA and with $D$ varying from
5\,cm$^{-1}$ to 60\,cm$^{-1}$. The choice $D=60$\,cm$^{-1}$ together with
the conversion factor  $\hbar^2/2\mu=8.7802375$\,cm$^{-1}${\AA}$^2$ 
(which  was used for all values of $D$) approximately represents the
interaction between the  Ar atom and the H$_2$ molecule \cite{connor}.\\

In Tables \ref{e8} and \ref{g8}  the energies and widths of the
first resonant states in the partial wave $\ell=8$ are presented 
for different values of $D$. The results obtained with
three other methods described in  Refs. \cite{connor,weon} are also given.
The digits shown there are stable under changes of the rotation angle 
and thus they indicate  the accuracy  achieved. The third column of
these tables, contains the results obtained in Ref. \cite{connor} using 
a Complex Rotation (CR) method which in some aspects is similar to ours. 
The authors of that reference  perform  the rotation directly in the 
Schr\"odinger equation and integrate it from $r=0$ outwards and from
a large $r_{\rm max}$ inwards. At the origin they use the WKB boundary
conditions and at $r_{\rm max}$ they start from the Siegert spherical wave.
In other words, the wave function is calculated using  physical boundary
conditions. In such an approach a resonance corresponds to a complex energy
which matches the inward and outward integration. As indicated  in
Ref. \cite{connor}, this method fails for broad resonances due to
instability in the outward integration.
In the fourth column  the results obtained in Ref. \cite{connor} using
the Quantum Time Delay (QTD) method are cited. This method is expected 
to be reliable for narrow resonances but its applicability to 
broad states is questionable. 
Finally, in the last column of Table \ref{e8} and \ref{g8} we give
the results obtained in Ref. \cite{weon} using the Finite Range Scattering
Wave (FRSW) method. The main idea of this method is based on the
fact that while the scattering wave function cannot be expanded properly
by a finite number of square integrable functions on an infinite range,
it is possible to do so for a finite range.\\

The test calculations show that our method works  well, especially for
narrow resonances. Broad resonances can also be located.
In contrast to the CR--method of Ref.\cite{connor}, which was unstable
for broad resonances corresponding to $D<35\,{\rm cm}^{-1}$,
we succeeded even in the case of $D=5\,{\rm cm}^{-1}$ which generates
an extremely broad state (its width is greater than the resonance energy by
a factor of 2). Our results for small values of $D$, reproduce well
the curve depicted in Fig. 3 of Ref. \cite{connor} which was  produced
semi-classically. The greater stability of the Jost function method as
compared to the CR--method of Ref. \cite{connor} can be attributed to
the use of the ansatz (\ref{ansatz}) for the regular solution. The
Riccati--Hankel functions, explicitly extracted there, describe correctly
all oscillations at large distances with the remaining functions
$F_\ell^{(\pm)}$ being smooth. Another reason for this
stability is the use of the  deformed integration path shown
on Fig. \ref{contour}, which enables us  to avoid fast oscillations
at short distances.
\section{Aziz potentials}
The  model potential considered  in the previous section, though
of typical form for inter-molecular interactions, does not describe any real
physical system. To give a more practical example, we apply
our formalism to study the interaction between two $^4$He atoms.
This interaction is of  interest in  the Bose--Einstein condensation and 
super-fluidity of helium at extremely low temperatures.  It is known that 
two helium atoms form a  dimmer molecule with binding  energy of 
$\sim 1$\,mK, but, to the best of  our knowledge, the possibility of 
forming  dimmer resonances has not  been investigated yet.\\

The search for a realistic $^4$He--$^4$He potential is a
long-standing problem in molecular physics. The earliest successful
potential of the Lennard--Jones(12-6) form was fitted just to reproduce
the second virial coefficient. Later on some other characteristics of
helium gas, such as viscosity, were included into the fitting (for 
a more detail review see Refs. \cite{aziz1,aziz2}). Nowadays,  
the potentials suggested by Aziz and co-workers are considered as be 
realistic.  Therefore,  in this
section, we apply our method using two versions of these potentials,
namely,  the HFDHE2 \cite{aziz1} and  the HFD-B \cite{aziz2} potentials. 
They can both be described using the same analytical form
\begin{eqnarray}
\label{azizpot}
      V(r)&=&\varepsilon\left[A\exp(-\alpha \zeta - \beta \zeta^2) -
      \left(\frac{C_6}{\zeta^6}+\frac{C_8}{\zeta^8}+
      \frac{C_{10}}{\zeta^{10}}\right)F(\zeta)\right]\ ,\\
\nonumber
      F(\zeta)&=&\left\{
\begin{array}{lcl}
      \displaystyle
      \exp\left[-\left({B}/{\zeta}-1\right)^2\right]\ &, &
      {\rm if}\ \zeta\le B\ ,\\
      1\ &, & {\rm if}\ \zeta>B\ ,\\
\end{array}
      \right.\\
\nonumber
      \zeta&=&r/r_m\ ,
\end{eqnarray}
but with different choices of the parameters (see Table \ref{azpar}). The
only principal difference in the functional form between them is the absence
of the Gaussian term ($\beta=0$) in the HFDHE2 potential.\\

Formally, the HFDHE2 and HFD-B  are regular potentials since  the
presence of the cut-off function $F(\zeta)$ in (\ref{azizpot}) makes
them finite at $r=0$,
\begin{equation}
\label{az0}
      V(r)\ \mathop{\longrightarrow}_{r\to 0}\ \varepsilon A\ .
\end{equation}
However, the product $\varepsilon A$ is very large ($\sim 10^6$) as compared
with the values of the potential in the attractive region. This
causes numerical instabilities when one tries to solve the Schr\"odinger
equation using  methods designed for regular potentials.
To avoid this difficulty, we notice that like in the case of singular
potentials the fast growth of the repulsion near the origin allows
the use  of the WKB boundary conditions near $r=0$. Indeed, the
derivative of the potential in the  vicinity of this point,
\begin{equation}
\label{paz0}
      \frac{{\rm d}V}{{\rm d}r}\ \mathop{\longrightarrow}_{r\to 0}\
      -\frac{\alpha\varepsilon A}{r_m}\ ,
\end{equation}
is of the same order of magnitude as $V$, which makes the derivative of the
local wavelength (\ref{crit1}) very small because of the large $A$,
\begin{equation}
\label{waz}
       |{\rm d}\lambda/{\rm d}r|\ \mathop{\longrightarrow}\limits_{r\to
       0}\ \frac{\alpha}{2r_m\sqrt{\displaystyle
       \frac{2\mu}{\hbar^2}\varepsilon A}}\ ,
\end{equation}
where the conversion factor $\hbar^2/2\mu=12.12$~K\AA$^2$, corresponding to
the choice of the units in Table \ref{azpar}, should be used. With the
parameters given, formula (\ref{waz}) gives 0.003 and 0.004 for
the potentials HFDHE2 and HFD-B respectively. These values of
$|{\rm d}\lambda/{\rm d}r|$ are small enough to comply with (\ref{wkbcond})
and allow the use of WKB boundary conditions. We can, therefore,
apply the method described in the preceding sections, to
the potentials HFDHE2 and HFD-B as if they were singular potentials.\\

To begin with, we tested the ability of our method to deal with this kind
of potentials by calculating the dimmer  binding energy. 
The results of these calculations are given in Table \ref{azbind} where, 
for comparison, we also cite the binding energies obtained in several 
earlier works. It is seen that the potentials HFDHE2 and HFD-B support 
a dimmer  bound state at energies which differ by a factor of 2. 

\begin{center}
\begin{figure}
\centerline{\epsfig{file=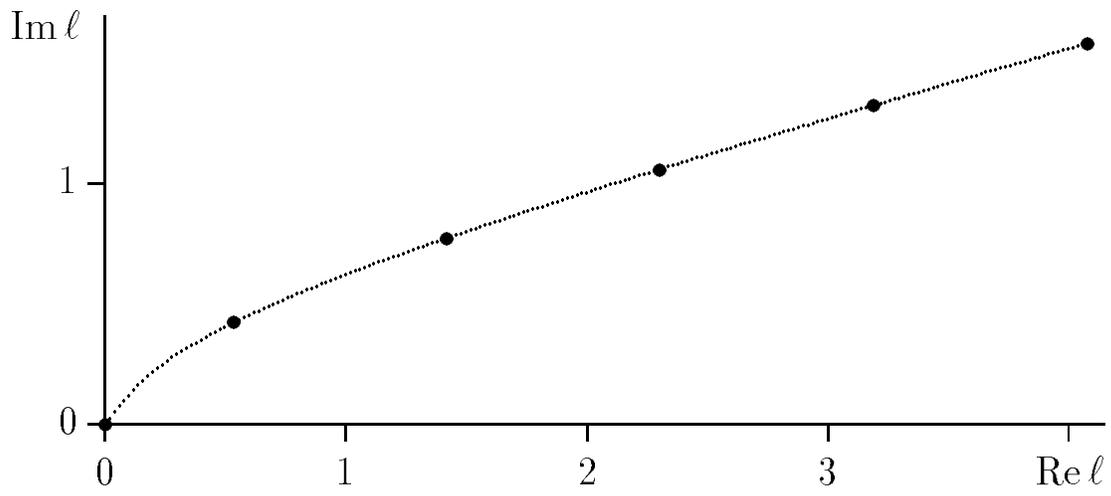,height=7.68362cm,width=16cm}}
\caption{Regge trajectory for the HFD-B potential. Filled circles indicate
 bound and resonant states.
} 
\label{azreggfig}
\end{figure}
\end{center}

A question then arises whether these potentials generate also quite 
different distribution of resonances which would result in different on 
and off the energy shell characteristics of the scattering amplitude. 
To study this we located several zeros of the Jost function in the
momentum as well as in the $\ell$--plane (Regge poles) for both potentials.
Due to the absence of a potential barrier there are no resonances in the
$S$--wave (at least with a reasonably small width). They appear, however,
at higher partial waves, starting from $\ell=1$. The energies and
widths of several such resonances are given in Table \ref{azrez}. They are
the lowest resonant states in each partial wave as they belong to the same
Regge trajectory which starts from the ground state. The trajectories for the
potentials HFDHE2 and HFD-B are practically indistinguishable and are shown
in Fig. \ref{azreggfig} by a single curve. Few  points of this curve
which correspond to resonances, are also given in Table \ref{azreggtab}.
It is seen that, to all practical purposes, the position of
the Regge poles are  the same. \\

As can be seen in Table \ref{azrez}, in each partial wave the 
potential generates a broad resonance  which covers the  whole 
low--energy region. This, together with the fact that the bound state 
pole of the amplitude is very close to  $k=0$, implies that the 
cross--section at  energies  $\sim 10$\,$^\circ$K ($\sim 10^{-3}$\,eV) 
must be quite large.
%
\section{Conclusions}
We presented an exact method for calculating the Jost solutions
and the Jost function for singular potentials, for real or complex momenta 
of physical interest. We demonstrated in the examples considered,
the suggested method is sufficiently stable and effective not only in 
the case of true singular potentials but also when a potential has 
strong, though finite, repulsion at short distances.\\

The method is based on simple differential equations of the first
order, which can be easily solved numerically. Thus, the spectrum generated
by any given potential can be thoroughly investigated.  At the same time,
physical wave function can be obtained having the correct asymptotic
behavior. When the potential has a Coulomb tail one  can simply replace the
Riccati--Hankel functions in the relevant equations by their Coulomb
analogous, $H_\ell^{(\pm)}(\eta,kr)\equiv F_\ell(\eta,kr)\mp iG_\ell(\eta,kr)$
 \cite{exact}. In the case of a non-central potential the Jost
function as well as the differential equations assume a matrix form with
somewhat more complicated, but still tractable  boundary conditions
at $r=0$  \cite{coupled}.\\

The method is also applicable when the angular momentum is complex. This
enables us to locate  Regge trajectories as well. This could be useful,
for example, in molecular scattering problems where the partial
wave series in many cases converges slowly \cite{connor1}.
This slow convergence can be overcome by allowing the angular momentum to
become complex valued which allows the  use of the Watson transformation. 
However, such a procedure   requires the knowledge of the positions of 
the Regge poles.\\

\section*{Acknowledgements}
Financial support from the  University of South Africa, the Foundation for
Research Development (FRD) of South Africa, and the Joint Institute for
Nuclear Research (JINR), Dubna, is greatly appreciated.\\


\newpage
\begin{table}
\begin{tabular}{|c|l|l|l|l|}
\hline
 Ref. & This work & CR\cite{connor} & QTD\cite{connor} & FRSW\cite{weon}\\
$D$ (cm$^{-1}$) & $E_{\rm res}$ (cm$^{-1}$) & $E_{\rm res}$ (cm$^{-1}$) &
$E_{\rm res}$ (cm$^{-1}$) & $E_{\rm res}$ (cm$^{-1}$) \\
\hline
5  & 30         &        &        &         \\
10 & 27         &        &        &         \\
15 & 25.5       &        &        &         \\
20 & 24.5       &        &        &         \\
25 & 22.90      &        &        &         \\
30 & 21.193     &        &        &         \\
35 & 19.450     & 19.449 & 19.370 &         \\
40 & 17.6478    & 17.647 & 17.619 & 17.617  \\
45 & 15.7768    & 15.777 & 15.769 & 15.769  \\
50 & 13.81980   & 13.820 & 13.819 & 13.818  \\
55 & 11.744242  & 11.744 & 11.744 & 11.743  \\
60 &  9.4943275 &  9.494 &  9.494 &  9.4934 \\
\hline
\end{tabular}
\vspace*{1cm}
\caption{Energies of the lowest resonances, in the $\ell=8$ partial wave,
for the Lennard--Jones potential with different $D$.}
\label{e8}
\end{table}
\begin{table}
\begin{tabular}{|c|l|l|l|l|}
\hline
 Ref. & This work & CR\cite{connor} & QTD\cite{connor} & FRSW\cite{weon}\\
$D$ (cm$^{-1}$) & $\Gamma_{\rm res}$ (cm$^{-1}$) &
$\Gamma_{\rm res}$ (cm$^{-1}$) &
$\Gamma_{\rm res}$ (cm$^{-1}$) & $\Gamma_{\rm res}$ (cm$^{-1}$) \\
\hline
5  & 60       &       &        &       \\
10 & 42       &       &        &       \\
15 & 29.4     &       &        &       \\
20 & 24.8     &       &        &       \\
25 & 18.63    &       &        &       \\
30 & 13.70    &       &        &       \\
35 & 9.724    & 9.727 & 10.228 &       \\
40 & 6.533    & 6.536 & 6.604  & 6.603 \\
45 & 4.039    & 4.039 & 3.992  & 3.990 \\
50 & 2.1833   & 2.183 & 2.143  & 2.142 \\
55 & 0.93915  & 0.939 & 0.926  & 0.926 \\
60 & 0.264474 & 0.264 & 0.263  & 0.263 \\
\hline
\end{tabular}
\vspace*{1cm}
\caption{Widths of the lowest resonances, in the  $\ell=8$ partial wave,
for the  Lennard--Jones potential with different $D$.}
\label{g8}
\end{table}
\begin{table}
\begin{tabular}{|c|l|l|}
\hline
parameter & HFDHE2 & HFD-B \\
\hline
$\varepsilon$ (K)    & 10.8             & 10.948 \\
$r_m        $ (\AA)  & 2.9673           & 2.963 \\
$A          $        & 544850.4         & 184431.01 \\
$\alpha     $        & 13.353384        & 10.43329537 \\
$\beta      $        & 0                & 2.27965105 \\
$C_6        $        & 1.3732412        & 1.36745214 \\
$C_8        $        & 0.4253785        & 0.42123807 \\
$C_{10}     $        & 0.178100         & 0.17473318 \\
$B          $        & 1.241314         & 1.4826 \\
\hline
\end{tabular}
\vspace*{1cm}
\caption{Parameters of the two versions of the Aziz $^4$He--$^4$He
potential.}
\label{azpar}
\end{table}
\begin{table}
\begin{tabular}{|c|l|l|}
\hline
&\multicolumn{2}{c|}{$^4$He-$^4$He binding energy (mK)}\\
\cline{2-3}
Ref. & HFDHE2 & HFD-B \\
\hline
This work    & 0.8301249029 & 1.6854110471 \\
\cite{shura} & 0.8301       & 1.6854       \\
\cite{aziz2} & --           & 1.684        \\
\cite{glock} & 0.830        & --           \\
\cite{lim}   & 0.829        & --           \\
\cite{uang}  & 0.8299       & --           \\
\hline
\end{tabular}
\vspace*{1cm}
\caption{Binding energies of $^4$He$_2$ di-atomic molecule for
the two versions of the Aziz potential.}
\label{azbind}
\end{table}
\begin{table}
\begin{tabular}{|c|r|r|r|r|}
\hline
&\multicolumn{2}{c|}{HFDHE2} & \multicolumn{2}{c|}{HFD-B} \\
\cline{2-5}
\phantom{+}$\ell$\phantom{+} & $E$ (K) & $\Gamma$ (K) &
$E$ (K) & $\Gamma$ (K) \\
\hline
1 & 0.334  & 1.822  & 0.339  & 1.795  \\
2 & 2.164  & 6.825  & 2.179  & 6.774  \\
3 & 5.930  & 15.195 & 5.963  & 15.098 \\
4 & 11.954 & 27.196 & 12.005 & 27.000 \\
5 & 20.478 & 42.949 & 20.518 & 42.603 \\
\hline
\end{tabular}
\vspace*{1cm}
\caption{Energies and widths of the lowest resonant states generated
by the two versions of the Aziz potential in several partial waves.}
\label{azrez}
\end{table}
\begin{table}
\begin{tabular}{|c|c||c|c|}
\hline
\multicolumn{2}{|c||}{HFDHE2} & \multicolumn{2}{c|}{HFD-B} \\
\hline
$E$ (K) & $\ell$ & $E$ (K) & $\ell$ \\
\hline
0.334  & $0.525+i0.429$ & 0.339  & $0.535+i0.424$ \\
2.164  & $1.408+i0.778$ & 2.179  & $1.416+i0.773$ \\
5.930  & $2.294+i1.065$ & 5.963  & $2.303+i1.060$ \\
11.954 & $3.182+i1.332$ & 12.005 & $3.191+i1.326$ \\
20.478 & $4.070+i1.590$ & 20.518 & $4.078+i1.584$ \\
\hline
\end{tabular}
\vspace*{1cm}
\caption{Regge poles corresponding to  resonances generated
by the two versions of the Aziz potential.}
\label{azreggtab}
\end{table}
\newpage
\end{document}